\makeatletter
\declare@file@substitution{revtex4-1.cls}{revtex4-2.cls}
\makeatother

\documentclass[]{aastex62}

\usepackage{amsmath}
\usepackage{mathrsfs}
\usepackage[caption=false]{subfig}
\usepackage{threeparttable}
\usepackage{booktabs}
\usepackage{longtable}

\received{27-May-2025}
\revised{07-Jun-2025}
\accepted{17-July-2025}

\shorttitle{Low-Redshift Excess in SN/GRBs}
\shortauthors{Li et al.}

\begin{document}

\title{Detection of Low-Redshift Excess in Supernova-Linked Gamma-Ray Bursts}

\correspondingauthor{Sheng-Bang Qian}
\email{qiansb@ynu.edu.cn}

\author{Qin-Mei Li}
\affiliation{Department of Astronomy, School of Physics and Astronomy, Key Laboratory of Astroparticle Physics of Yunnan Province, Yunnan University, Kunming 650091, China}

\author{Qi-Bin Sun}
\affiliation{Department of Astronomy, School of Physics and Astronomy, Key Laboratory of Astroparticle Physics of Yunnan Province, Yunnan University, Kunming 650091, China}

\author{Sheng-Bang Qian}
\affiliation{Department of Astronomy, School of Physics and Astronomy, Key Laboratory of Astroparticle Physics of Yunnan Province, Yunnan University, Kunming 650091, China}

\author{Fu-Xing Li}

\affiliation{Department of Astronomy, School of Physics and Astronomy, Key Laboratory of Astroparticle Physics of Yunnan Province, Yunnan University, Kunming 650091, China}

\begin{abstract}

Gamma-ray bursts (GRBs) are traditionally classified into long (lGRBs) and short (sGRBs) durations based on their $T_{90}$, with lGRBs widely used as tracers of the cosmic star formation rate (SFR) due to their observed association with core-collapse supernovae. However, recent detections of kilonovae accompanying some lGRBs challenge this assumption, suggesting potential contamination from compact binary mergers. Here, we move beyond the conventional $T_{90}$-based classification and focus exclusively on GRBs directly associated with supernovae — the most direct signatures of massive stellar collapse — to reassess their connection to the SFR. Using a sample of SN/GRBs, we construct the luminosity–redshift ($L$–$z$) plane and uncover a significant correlation between these variables. To account for observational biases, we apply the $\tau$ statistic and Lynden-Bell’s $C^{-}$ method to derive the intrinsic luminosity function and formation rate. Our analysis reveals that even among this well-defined subsample, the SN/GRB formation rate still exceeds the SFR at low redshifts ($z < 1$). These findings suggest that GRBs at low redshift may not serve as reliable tracers of the SFR, and that larger samples are required to further investigate this discrepancy.

\end{abstract}


\section{Introduction} \label{sec:intro}
Gamma-ray bursts (GRBs) are among the most energetic astrophysical phenomena in the universe, characterized by intense emissions of high-energy photons. 
Their brightness at high redshifts surpasses that of supernovae (SNe), making GRBs particularly useful for investigating various cosmological phenomena, such as dark energy \citep{2011ApJ...727L..34W,2023MNRAS.521.4406L,2023MNRAS.518.2201D}, metal enrichment \citep{2012ApJ...760...27W,2016IAUFM..29B.261C}, star formation rate (SFR; \citealp{2015ApJS..218...13Y,2024MNRAS.527.7111L,2024ApJ...963L..12P}). Notably, GRB 090429B holds the record as the most distant detected burst at $z=9.4$ \citep{2011ApJ...736....7C}.

GRBs have traditionally been classified into two categories based on their duration $T_{90}$ \citep{1993ApJ...413L.101K}. Short GRBs (sGRBs; $T_{90} < 2\,\text{s}$) are generally originated from the mergers of compact binary systems, such as neutron star-black hole (NS-BH) or binary neutron stars (BNS). The detection of GW170817 and its electromagnetic counterpart sGRB 170817A provided definitive evidence that BNS mergers can give rise to sGRBs \citep{2017ApJ...848L..13A}. This event is often accompanied by kilonova (KN), which is optical transients powered by r-process nucleosynthesis \citep{2021ApJ...916...89R}.

Long GRBs (lGRBs; $T_{90} > 2\,\text{s}$) are widely believed to originate from the collapse of massive stars. This association is supported by multiple lines of evidence: (1) direct observational links between lGRBs and supernovae (SNe) \citep{2003ApJ...591L..17S,2004ApJ...609..952Z}, and (2) their preference for occurring in star-forming galaxies \citep{2002AJ....123.1111B} and regions \citep{2006Natur.441..463F}. Consequently, the formation rate (FR) of lGRBs has been expected to trace the cosmic star formation rate (SFR) \citep{2012A&A...539A.113E}. Understanding the precise relationship between the lGRB rate and the cosmic SFR is therefore essential for using lGRBs as
cosmological probes.

However, a growing body of evidence suggests deviations from this expected correlation. \citet{2015ApJS..218...13Y} first reported an excess of lGRBs at low redshifts ($z<1$) based on a sample of 127 lGRBs, taking into account differential comoving volume effects and luminosity-redshift co-evolution. Notably, their sample included KN/GRB 060614 and KN/GRB 070714B, now considered potential products of compact binary mergers. They suggested that this discrepancy might stem from ambiguities in defining lGRBs. This concern was later echoed by \citet{2020ApJ...902...40Z}, who proposed a revised duration threshold of 1s based on \textit{Swift} observations.

While the lGRB-SN connection provides strong support for the collapsar model, several notable exceptions challenge this paradigm. For instance, lGRB 060614 \citep{2006Natur.444.1044G} and lGRB 060505 \citep{2006Natur.444.1047F} show no evidence of an accompanying SN but instead exhibit KN signatures. Recent work by \citet{2024ApJ...963L..12P} quantifies this divergence, showing that the lGRBs FR exceeds the cosmic SFR at $z<2$, with approximately 60\% of lGRBs potentially originating from compact object mergers and only 40\% from collapsars. Similarly, sGRB 200826A ($T_{90}=1.16$\,s) has been observed to be associated with a SN \citep{2021NatAs...5..911Z}. In addition, several lGRBs have recently been confirmed to be linked with KN, including GRB 060505 \citep{2021arXiv210907694J}, GRB 080503A \citep{2023ApJ...943..104Z}, GRB 211211A \citep{2022Natur.612..232Y}, and GRB 230307A \citep{2023ApJ...953L...8W}, all of which are thought to arise from binary compact object mergers.

These findings collectively suggest that a significant fraction of lGRBs may not originate from core-collapse supernovae, but rather from compact binary mergers. As a result, the traditional association between lGRBs and massive star collapse is no longer universally valid. Similarly, the link between sGRBs and compact binary mergers is also being reconsidered. This undermines the reliability of using $T_{90}$-based lGRBs as
tracers of the cosmic SFR. To address this issue, this paper proposes that GRBs directly associated with supernovae (SN/GRBs) should be used as more reliable tracers of the cosmic SFR. By focusing on this subset of GRBs, we can obtain a cleaner progenitor population, minimizing contamination from compact binary merger events. This approach offers a more robust foundation for utilizing GRBs in cosmological studies.

 This paper is organized as follow. In Section 2, we introduce sample selection and data analysis method. In Section 3, we show the result of luminosity function (LF) and FR for 56 SN/GRBs. The conclusion and discussion are displayed in Section 4. Throughout the paper, we assume a flat $\Lambda $ universe with ${\Omega _m} =0.3$ and ${H_0} =$ 70 $km\,{s^{ - 1}}Mp{c^{ - 1}}$.

\section{Sample Selection and Data Analysis method} \label{sec:sample}
\subsection{Sample Selection}
 \citet{2023MNRAS.524.1096L} systematically compiled a sample containing 53 SN/GRBs and 15 kilonova-associated GRBs (KN/GRBs). Subsequent observations have further expanded this sample: \citet{2021GCN.29306....1R} identified supernova features in the Large Binocular Telescope (LBT) spectra of GRB 201015A at redshift $z=0.42$. \citet{2022GCN.32800....1D} confirmed supernova signatures in GRB 221009A through continuum subtraction analysis using OSIRIS on the 10.4-m Gran Telescopio Canarias (GTC). \citet{2021GCN.31098....1B} reported SN/GRB 211023A from optical observations. Additional cases include GRB 221009A \citep{2023ApJ...946L..22F} and GRB 230812B \citep{2024ApJ...960L..18S}. Combining main spectral data from \citet{2023MNRAS.524.1096L} and the Gamma-ray Burst Coordinates Network\footnote{\url{https://www.mpe.mpg.de/~jcg/grbgen.html}}, we establish a final sample of 58 SN/GRBs. GRB 071112C and GRB 150518A were excluded due to insufficient spectral data for analysis. After quality control, 56 well-observed SN/GRBs were used to determine the LF and FR. In Table \ref{tab:1}, we list the GRB sample, including name (Column 1),  duration $T_{90}$ (Column 2), redshift (Column 3), low-energy power-law index $\alpha$ (Column 4), high-energy power-law index $\beta$ (Column 5), peak energy $E_p$ of the $vfv$
 spectrum in the observer’s frame (Column 6), peak flux P in a certain energy range (Column 7), energy band (Column 8), bolometric luminosity (Column 9), and references (Column 10) of SN/GRBs. Figure~\ref{fig:1} (a) displays the $T_{90}$ duration distribution for our SN/GRBs sample. Notably, one exhibit durations shorter than the conventional 2\,s threshold. This observation, combined with the known population of long-duration ($T_{90} > 2\,\text{s}$) events originating from compact binary mergers (as discussed in previous sections), further demonstrates the limitations of using $T_{90}$ as a definitive progenitor classification criterion.  
 
Two spectral models are employed to fit the spectra of GRBs: a power law with an exponential cutoff model (CPL; \citealp{2008ApJS..175..179S}) and the Band model \citep{1993ApJ...413..281B}. The functional forms of these models are as follows:
	
For the CPL model:
\begin{equation}
	f(E) = A \left( \frac{E}{100\, \text{keV}} \right)^{\alpha} \exp \left( -\frac{(2 + \alpha) E}{E_p} \right),
	\tag{1}
\end{equation}
where $A$ is the normalization factor, $\alpha$ is the power-law index, and  $E_{\rm p}$ is the peak energy in observer frame.

and for the Band model:
\begin{equation}
	f(E) = 
	\begin{cases}
		A \left( \frac{E}{100~\mathrm{keV}} \right)^{\alpha} \exp \left( -\frac{(2 + \alpha) E}{E_{\rm p}} \right), & E < \frac{(\alpha - \beta) E_{\rm p}}{2 + \alpha}, \\
		A \left( \frac{E}{100~\mathrm{keV}} \right)^{\beta} \exp \left[ (\beta - \alpha) \left( \frac{(\alpha - \beta) E_{\rm p}}{(2 + \alpha) 100~\mathrm{keV}} \right)^{\alpha - \beta} \right], & E \geq \frac{(\alpha - \beta) E_{\rm p}}{2 + \alpha},
	\end{cases}
	\tag{2}
\end{equation}
where $A$ is the normalization factor, $\alpha$ and $\beta$ are the low- and high-energy photon indices, respectively.

Since the peak fluxes of GRBs are observed over a wide range of redshifts, corresponding to different rest-frame energy bands, we need to make K-correction transform the observed band of telescope in to  1-$10^4$ keV band to obtain the bolometric luminosity of GRBs \citep{2001AJ....121.2879B}. The bolometric luminosity $L$ of a GRB is given by:
\begin{equation}
	L = 4 \pi d_L^2(z) F K,
	\tag{3}
\end{equation}
where $d_L(z)$ is the luminosity distance at redshift $z$, defined as:
\begin{equation}
	d_L(z) = \frac{c}{H_0}(1+z) \int_0^z \frac{dz'}{\sqrt{1 - \Omega_m + \Omega_m (1 + z')^3}},
	\tag{4}
\end{equation}
$F$ is the peak flux observed between certain energy ranges ($E_{\rm min}$, $E_{\rm max}$), and $K$ is the K-correction factor. If the $F$ is in units of erg cm$^{-2}$ s$^{-1}$, then the parameter $K$ is defined as:
\begin{equation}
	K = \frac{\int_{1~\mathrm{keV}/(1+z)}^{10^4~\mathrm{keV}/(1+z)} E f(E) \, dE}{\int_{E_{\rm min}}^{E_{\rm max}} E f(E) \, dE}.
	\tag{5}
\end{equation}
If the flux $F$ is in units of photons cm$^{-2}$ s$^{-1}$, then the parameter $K$ is defined as:
\begin{equation}
	K = \frac{\int_{1~\mathrm{keV}/(1+z)}^{10^4~\mathrm{keV}/(1+z)} E f(E) \, dE}{\int_{E_{\rm min}}^{E_{\rm max}} f(E) \, dE}.
	\tag{6}
\end{equation}
Here, $f(E)$ represents the spectral model of GRBs.

It should be noted that the bursts in our sample were detected by instruments aboard different satellites. The flux sensitivities of these instruments are as follows: approximately $4 \times 10^{-8} \, \text{erg cm}^{-2} \text{s}^{-1}$ for CGRO/BATSE; $\sim 10^{-7} \, \text{erg cm}^{-2} \text{s}^{-1}$ for Konus-Wind, BeppoSAX, and Fermi/GBM; $\sim 3 \times 10^{-8} \, \text{erg cm}^{-2} \text{s}^{-1}$ for HETE-2; and about $10^{-8} \, \text{erg cm}^{-2} \text{s}^{-1}$ for Swift/BAT. Notably, some bursts were detected by more than one satellite. To ensure that all bursts in the sample are above the flux limit, we adopt the best sensitivity of $F_{\rm limit} = 1 \times 10^{-8} \, \text{erg cm}^{-2} \text{s}^{-1}$ for SN/GRBs as the flux threshold for the entire sample, following \citet{2012MNRAS.423.2627W}. Hence, the corresponding luminosity limit at redshift $z$ can be calculated using ${L_{\rm limit}} = 4\pi d_L^2(z) F_{\rm limit}$.

\subsection{Lynden-Bell's ${c^ - }$ method} \label{sec:c- method}


Several selection effects influence the observed redshift distribution of GRBs \citep{2007NewAR..51..539C}, and consequently their inferred event rate. 
The most significant of these arises from the observational limitations of satellite. These satellite has a flux limit, meaning that it is unable to detect GRBs fainter than a given threshold. As a result, the observed sample is truncated, and the intrinsic redshift distribution of GRBs cannot be reliably determined without first accounting for these observational biases.

\citet{1971MNRAS.155...95L} (Lynden-Bell's ${c^ - }$) putted forward a non-parametric methods that can be used to study the LF and density evolution for quasar sample. But \citet{1992scma.conf..173P} pointed that the drawback of non-parametric method is \textit{ad hoc} assumption of \textit{uncorrelated variables}. To overcoming this defects, \citet{1992ApJ...399..345E} (EP) developed a new method (EP-L method) to test whether luminosity (L) and redshift (z) are correlated or not. This method is used on a variety of cosmological objects, such as quasar \citep{1971MNRAS.155...95L,1992ApJ...399..345E}, Active galactic nucleus \citep{1999ApJ...518...32M,2014ApJ...786..109S,2021ApJ...913..120Z}, GRBs \citep{2015ApJS..218...13Y,2015ApJ...806...44P,2025ApJ...978..160L}, and FRB \citep{2019JHEAp..23....1D,2024ApJ...973L..54C}. In this paper, we also use this method to drive the cosmological evolution of LF and FR of our sample.

\begin{figure*}[ht!]  
	\centering
		\gridline{\fig{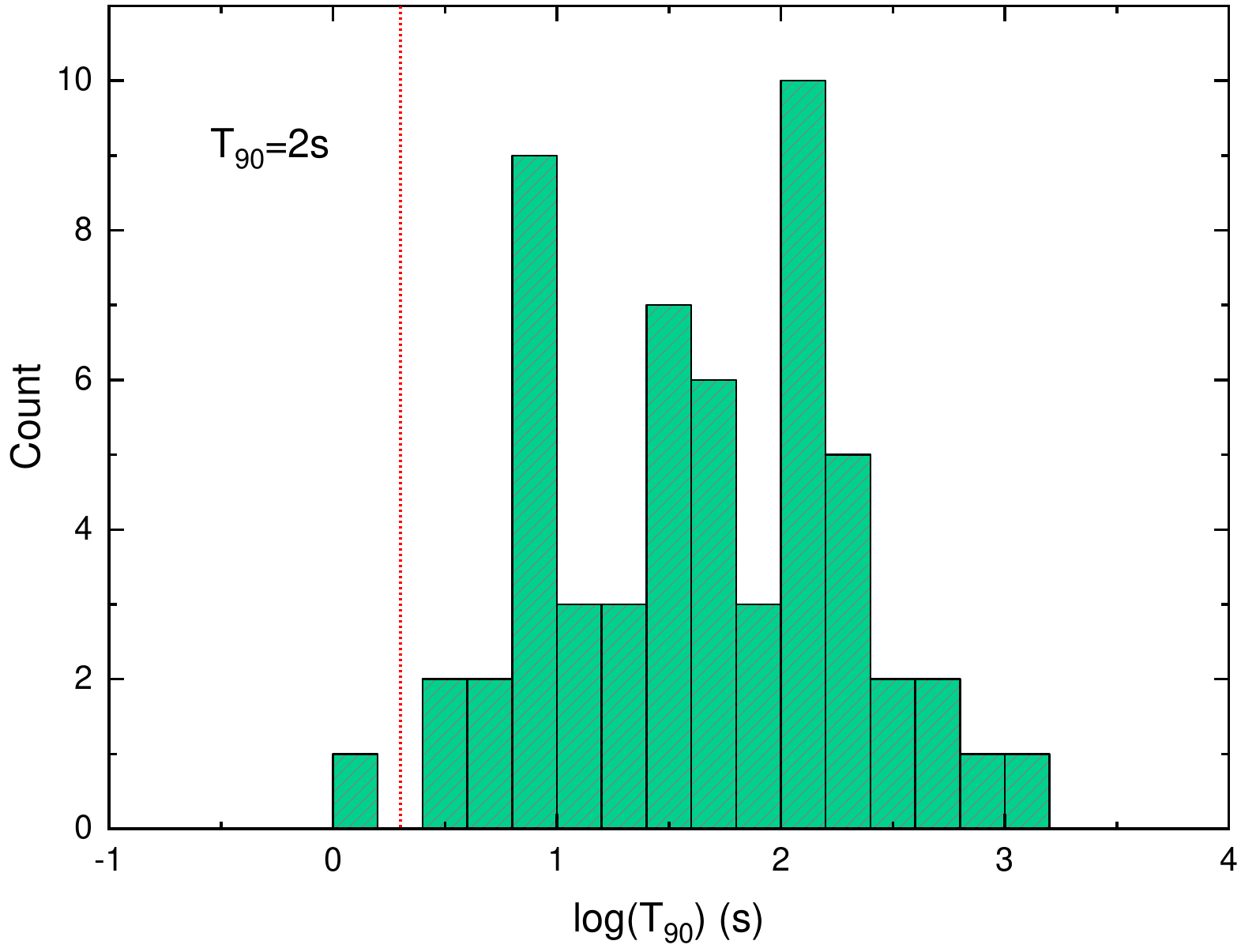}{0.45\textwidth}{(a)}
	\fig{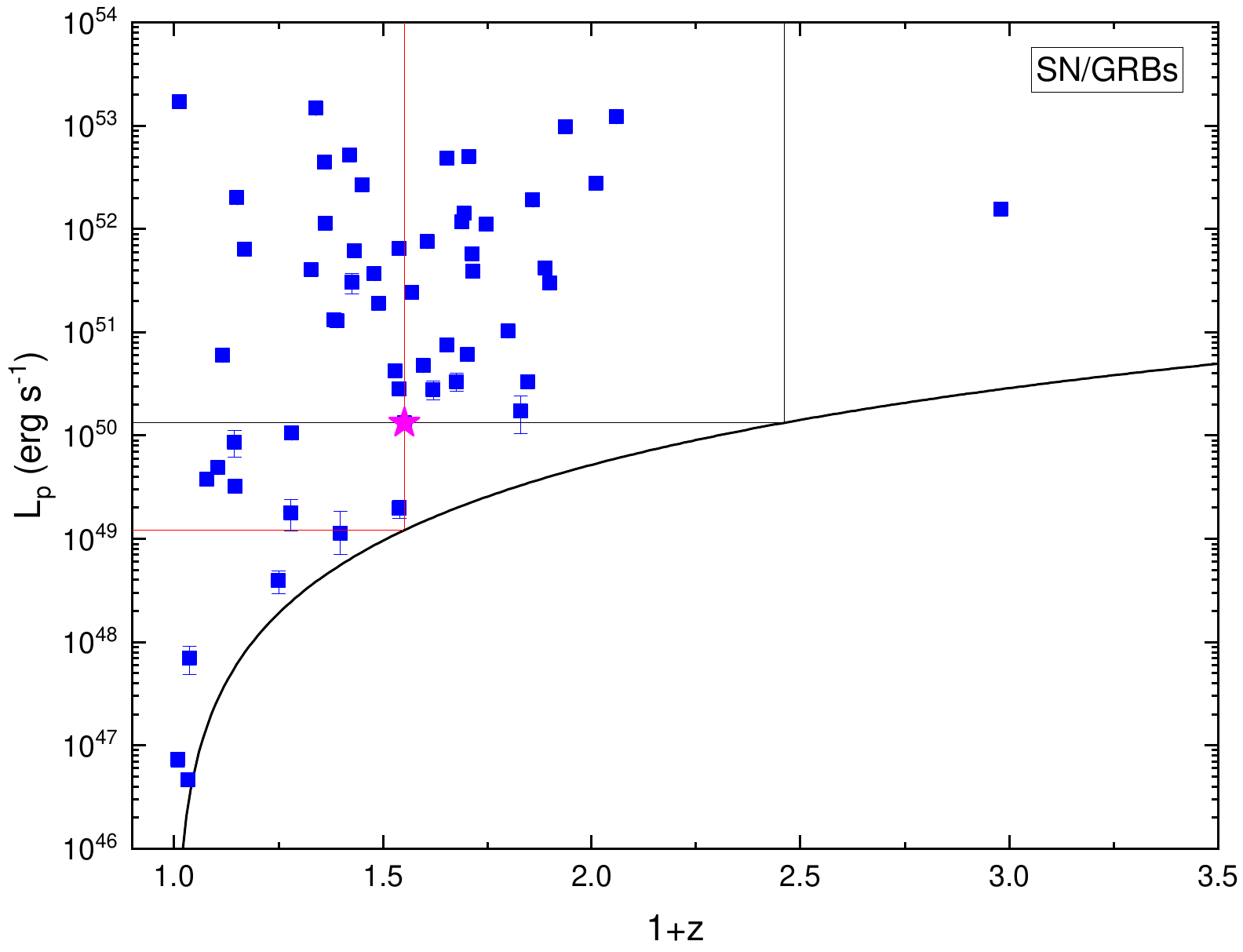}{0.45\textwidth}{(b)}
}
\gridline{\fig{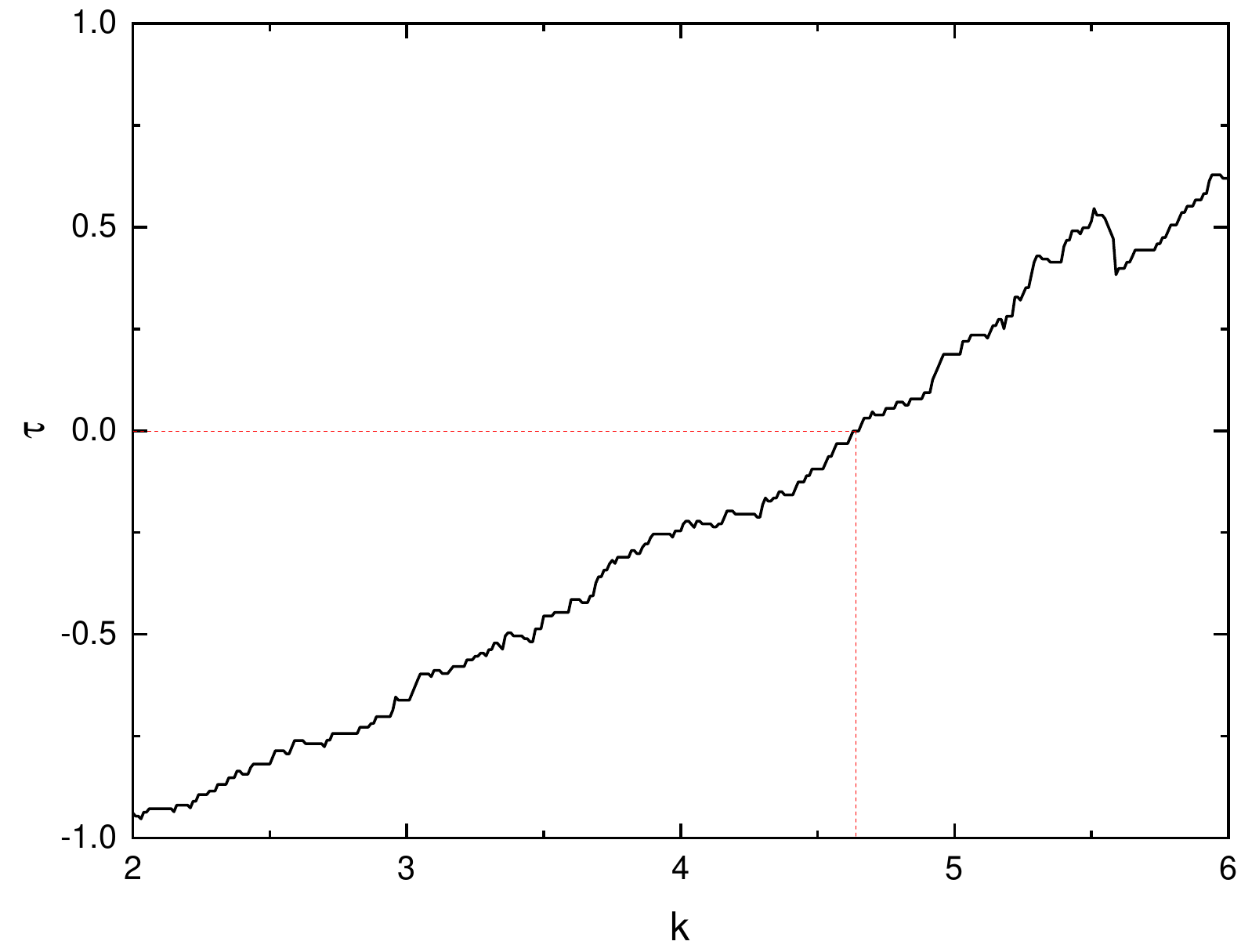}{0.45\textwidth}{(c)}
	\fig{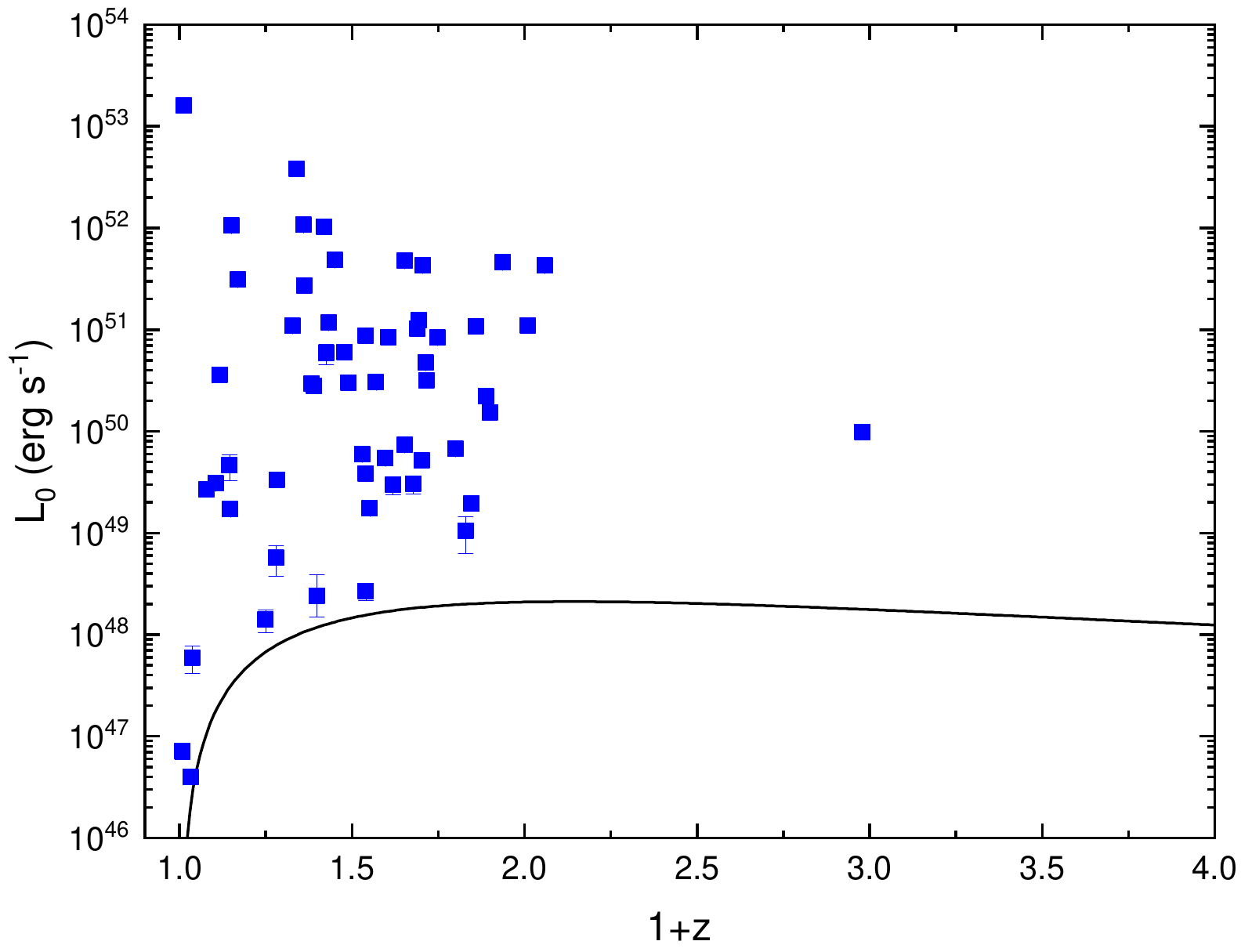}{0.45\textwidth}{(d)}
}			
	\caption{
	Distributions and correlations for the SN/GRBs sample: 
	(a) Duration ($T_{90}$) distribution of SN/GRBs events; 
	(b) Bolometric luminosity distribution where individual points represent different SN/GRBs, with the line indicating the sensitivity limit of $1.0 \times 10^{-8}\,\mathrm{erg\,cm^{-2}\,s^{-1}}$; 
	(c) In the Kendall $\tau$ correlation test, the red dotted line represents the null hypothesis ($\tau = 0$), and the measured correlation strength of $k = 4.64$ suggests that the evolutionary dependence between luminosity and redshift has been effectively removed; 
	(d) De-evolved luminosity function following $L = L_{0}(1 + z)^{4.64}$ for our sample of 56 SN/GRBs, removing the redshift evolution component.}
  \label{fig:1}
\end{figure*}

In Figure \ref{fig:1} (b), it is easy see that luminosity and redshift are dependent. Therefore, we need to eliminate this effect at first by using EP-L method \citep{1992ApJ...399..345E}. Luminosity evolves with function ${g_k}(z) = {(1 + z)^k}$, we can determine the value of $k$. As pointed as previous author \citep{1992ApJ...399..345E,2015ApJS..218...13Y,2019JHEAp..23....1D}, we use the kendall $\tau $ statistical method to drive the $k$.
Figure \ref{fig:1} (b) show the distribution of luminosity and redshift, for a random point $i$ (${z_i},L{}_i $) in this plane, the ${J_{\rm{i}}}$ can be defined as
\begin{equation}
{J_{\rm{i}}} = \{ {\rm{j}}|{L_{\rm{j}}} \ge {L_i},{z_j} \le z_i^{\max }\}
\end{equation}
where $L_i$ is the luminosity of $i$th SN/GRB and $z_i^{\max }$ is the maximum redshift at which the SN/GRBs with the luminosity limit. The range of ${J_{\rm{i}}}$ is shown as the black rectangle in Figure \ref{fig:1} (b). The number of SN/GRBs that contained in this region is defined as ${n_i}$. The number ${N_i}={n_i-1}$, which means takeing the $i$th point out. Similarly, the $J_i^{'}$ can be defined as
\begin{equation}
J_i^{'} = \{ {\rm{j}}|{L_{\rm{j}}} \ge L_{_i}^{\lim },{z_j} \le {z_i}\}
\end{equation}
where $L_{_i}^{\lim }$ is the limiting corrected gamma-ray luminosity at redshift ${z_i}$. In Figure \ref{fig:1} (b), the range of $J_i^{'}$ is shown as red rectangle. The number of SN/GRBs that contained in this range is defined as ${M_i}$.

Firstly, we consider to define the number of SN/GRBs that have redshift $z$ less than or equal to ${z_i}$ in black rectangle as ${R_i}$. The statistic $\tau$ is \citep{1992ApJ...399..345E}
\begin{equation}
\tau  = \frac{{\sum\nolimits_i {({R_{\rm{i}}} - {E_i})} }}{{\sqrt {\sum\nolimits_i {{V_{\rm{i}}}} } }}
\end{equation}
where ${E_i} = \frac{{1 + ni}}{2}$ and ${V_i} = \frac{{n_i^2 - 1}}{{12}}$ are the expected mean and the variance for the hypothesis of independence, respectively. 
If $R_i$ is exactly uniformly distributed between 1 and $n_i$, then the samples with $R_i \leq E_i$ and $R_i \geq E_i$ should be approximately equal in number, and the test statistic $\tau$ will be close to zero. If we choose a functional form of $g(z) = {(1 + z)^k}$ that make the test statistic $\tau = 0$, then the effect of luminosity evolution can be removed by applying the transformation $L_0 = L / g(z)$. Figure \ref{fig:1} (d) show the distribution of non-evolving gamma-ray luminosity. The $k$ value is 4.64 for 56 SN/GRBs, as shown in Figure \ref{fig:1} (c) .

After eliminate the effect of the luminosity evolution, we can obtain the cumulative luminosity function with non-parametric method from following equation
\begin{equation}
\psi ({L_{0{\rm{i}}}}{\rm{) = }}\mathop \prod \limits_{{\rm{j}} < i} (1 + \frac{1}{{{N_j}}})    
 \label{equ:5}
\end{equation}
and the cumulative number distribution can be derived from
\begin{equation}
\phi (z{\rm{) = }}\mathop \prod \limits_{j < i} (1 + \frac{1}{{{M_j}}})           
 \label{equ:6}
\end{equation}

The differential form of $\phi (z)$ represents the formation rate of $\rho $, which can be expressed as
\begin{equation}
\rho (z) = \frac{{d\phi (z)}}{{dz}}(1 + z){\left( {\frac{{dV(z)}}{{dz}}} \right)^{ - 1}}           
\label{equ:7}
\end{equation}
where ${\frac{{dV(z)}}{{dz}}}$ represents the differential comoving volume, which can be written as
\begin{eqnarray}
\begin{split}
\frac{{dV(z)}}{{dz}} =4\pi \left( {\left. {\frac{c}{{{H_0}}}} \right)} \right.\frac{{D_L^2}}{{{{(1 + z)}^2}}} \times \frac{1}{{\sqrt {1 - {\Omega _m} + {\Omega _m}{{(1 + z)}^3}} }} 
 \label{equ:8}
 \end{split}
\end{eqnarray}

\section{Result} 
\label{sec:luminosity function and formation rate of SN/GRB}
\begin{figure}[ht!]  
	\centering
	\includegraphics[width=3.5in]{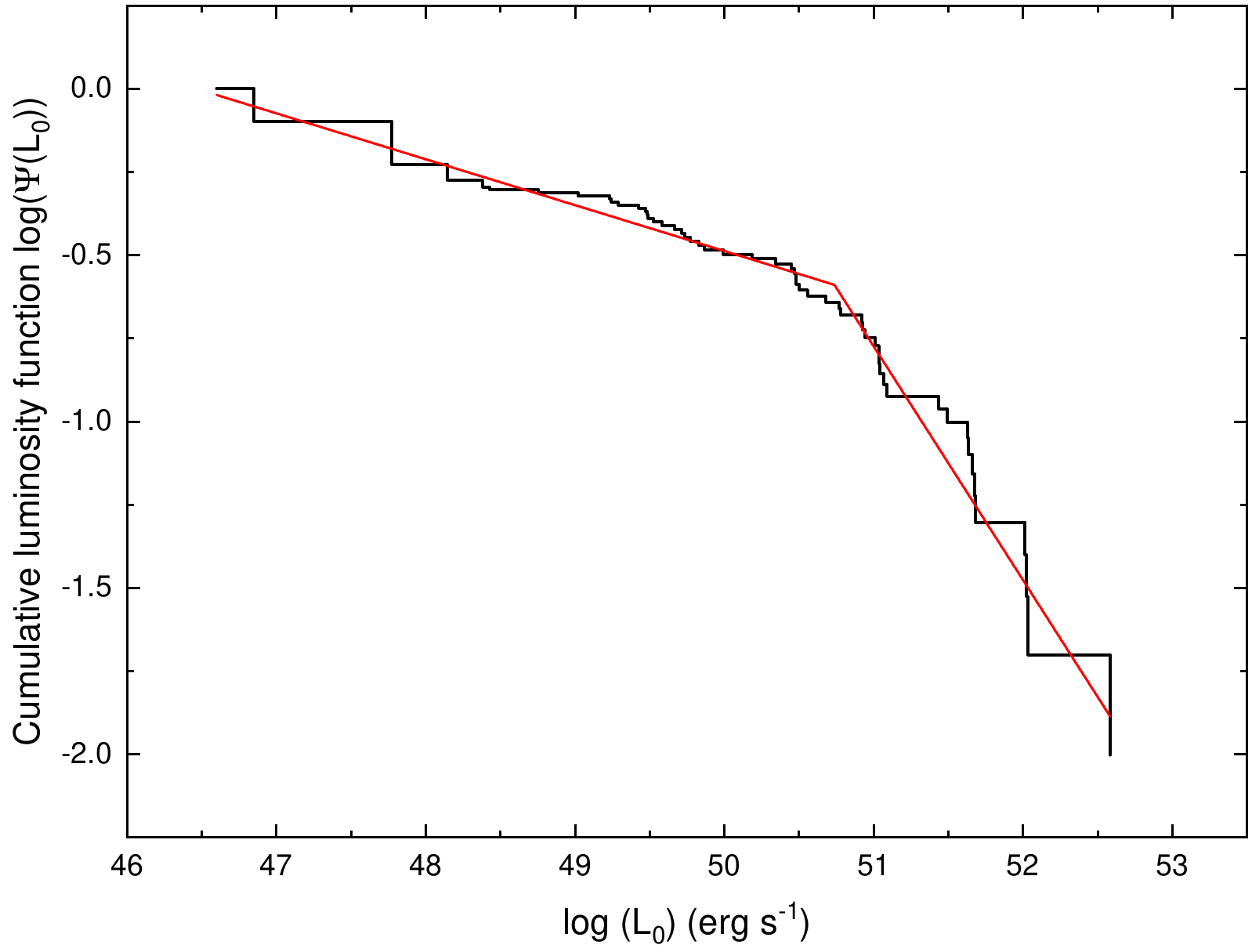}
	\includegraphics[width=3.5in]{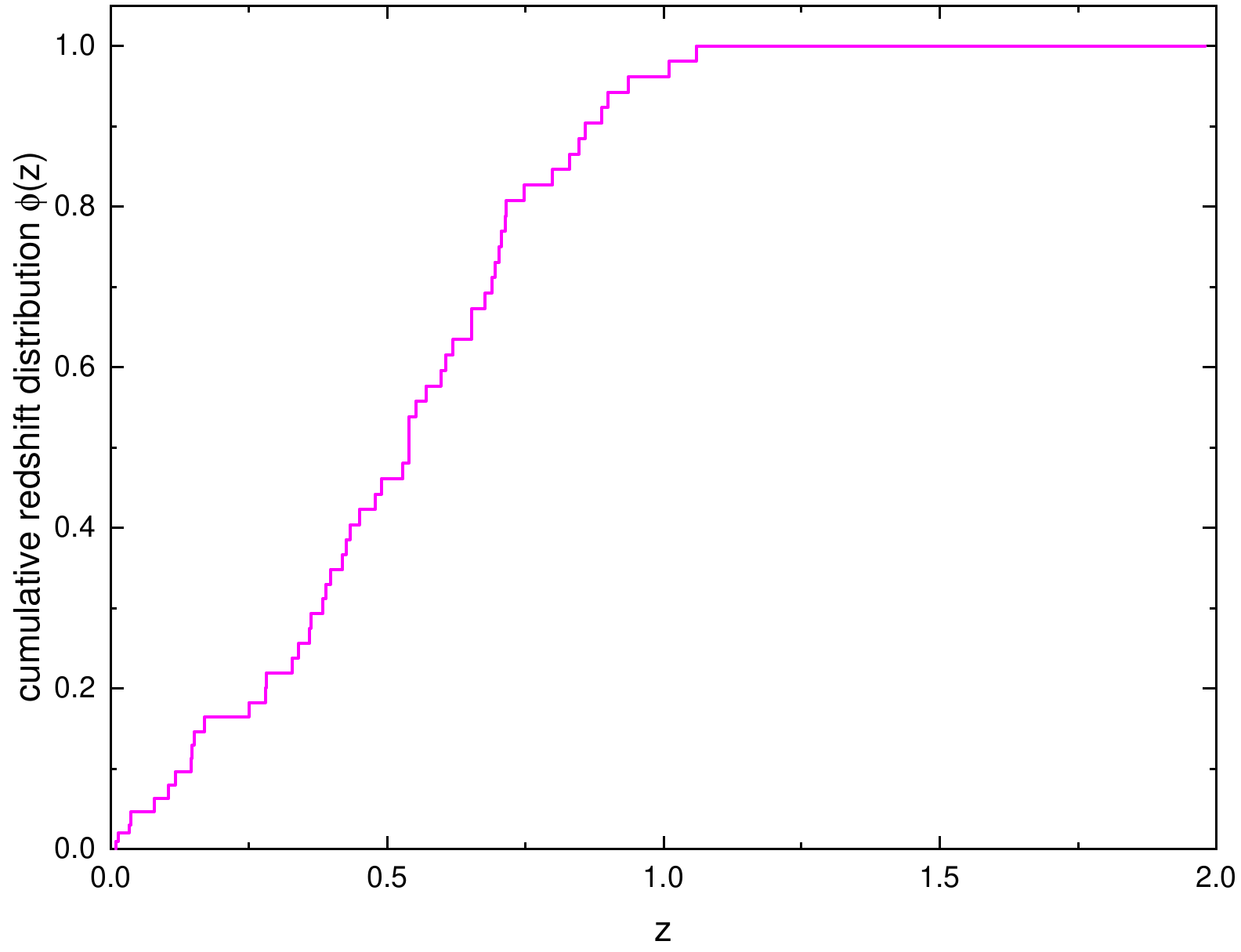}
	\caption{Left: The distribution of cumulative luminosity function $\psi({L_0})$, which is normalized to unity at the lowest luminosity. The function is fitted by red solid line with broken power law. The luminosity function can be expressed as $\psi ({L_0}) = L_0^{-0.14 \pm 0.01},{L_0} < L_0^b$ and $\psi ({L_0}) = L_0^{-0.70 \pm 0.03},{L_0} > L_0^b$; Right: Normalized cumulative redshift distribution.}
  \label{fig:2}
\end{figure}

In this section, we present the derived LF and FR for our SN/GRBs sample. Following the methodology described in previous sections, we quantify the luminosity evolution through a non-parametric approach using the Kendall $\tau$ correlation test.
We use broken power law to fit the  cumulative luminosity function in Figure \ref{fig:2} (a), and the following results were obtained:
\begin{eqnarray}
\psi ({L_0}) \propto \left\{ {\begin{array}{*{20}{c}}
{L_0^{ -0.14 \pm 0.01},{L_0} < L_0^b}\\
{L_0^{-0.70 \pm 0.03},{L_0} > L_0^b}                    
\end{array}} \right.
 \label{equ:9}
\end{eqnarray}\\
where $L_0^b = (5.50 \pm 0.67) \times 10^{50} \, \text{erg}\, \text{s}^{-1}$ is the break point. It is worth to point out that the $\psi (L/g(z))$ is local luminosity at $z=0$ for the luminosity evolution is removed, and the luminosity function can be rewrite as $\psi ({L_0}) = \psi {(L/(1 + z)^{4.64})}$. Therefore, the break luminosity at z can be deduced $L_z^b = L_0^b{(1 + z)^{4.64}}$.

\begin{figure}[ht!]  
	\includegraphics[width=5in]{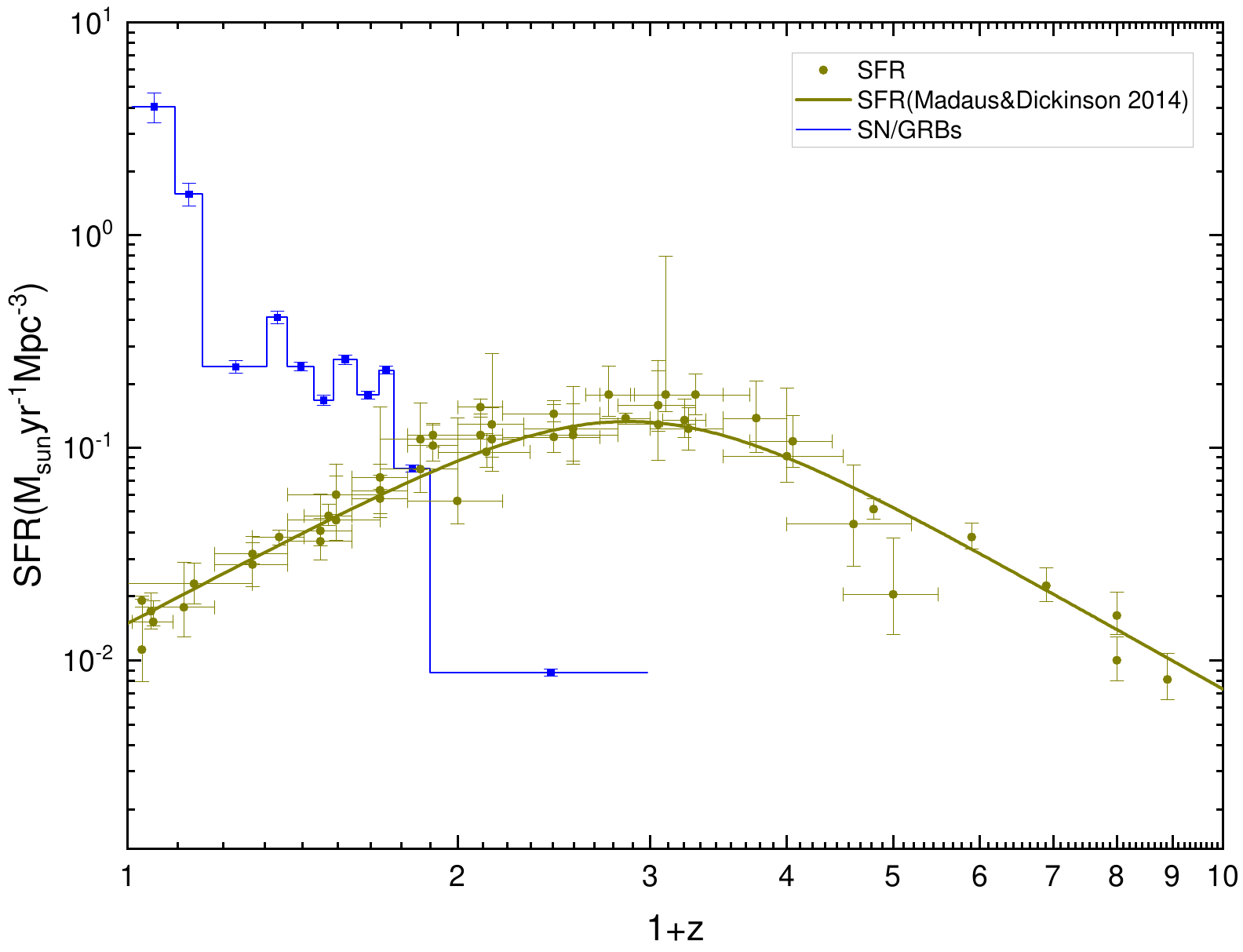}
	\centering
	\caption{Comparison SN/GRBs rate and SFR. The SFR data and fit line were collected from \citet {2014ARA&A..52..415M} (Brown dots). The blue line are the formation rate of fitting 56 SN/GRBs sources. The error bar gives a 1 $\sigma$ poisson error.}
  \label{fig:5}
\end{figure}

Figure~\ref{fig:2} (b) displays the cumulative redshift distribution $\phi(z)$ for our sample. Prior to deriving the FR, we first obtain the differential form $d\phi(z)/dz$ through numerical differentiation. According to Equation \ref{equ:7}, we can calculate the FR between two adjacent redshift bins. However, this approach leads to significant data fluctuations. To mitigate this issue, we instead compute the FR every five redshift points and assign the resulting value to the corresponding redshift interval. The error bar gives a 1 $\sigma$ poisson error \citep{1986ApJ...303..336G}. As evident in Figure~\ref{fig:5}, the resulting FR $\rho(z)$ exhibits a pronounced decreasing trend with increasing redshift. When compared to the cosmic SFR, our results reveal a persistent excess at low redshifts ($z< 1$).


\section{DISCUSSIONS AND CONCLUSIONS }

	In this work, we compile a sample of 58 GRBs that have been securely associated with core-collapse SN from the literature. Utilizing Lynden-Bell's $C^{-}$ method to eliminate the redshift-luminosity correlation, we derive the intrinsic luminosity function and formation rate of 56 SN/GRBs. This enables us to perform a robust comparison between the GRB event rate and the cosmic SFR. Our key findings are as follows:
(1) A Kendall $\tau$ test yields a statistically significant correlation with a $k$-value of 4.64 for the 56 SN/GRBs in our sample;
(2) We observe a notable excess in the SN/GRB event rate compared to the SFR at low redshifts ($z < 1$).
These results suggest that the observed low-redshift excess is unlikely to be fully explained by contamination from compact binary mergers, and may instead point to intrinsic differences in the GRB progenitor population or its dependence on environmental factors.

It is well established that GRBs can originate from either the mergers of compact binaries \citep{1986ApJ...308L..43P} or the core collapse of massive stars \citep{1993ApJ...405..273W}. Since these compact remnants form shortly after the birth of their massive stellar progenitors, GRBs are expected to trace the cosmic SFR \citep{2000ApJ...536....1L, 2000MNRAS.312L..35B}. However, due to orbital decay timescales, a time delay between the SFR and the merger rate is expected for compact binary systems \citep{1992ApJ...389L..45P, 2015MNRAS.448.3026W}. sGRBs are confirmed to arise from compact binary mergers, exhibit a rate distribution determined by the convolution of the SFR with the merging time delay distribution $P_m(\tau)$ \citep{2006A&A...453..823G}. Therefore, sGRBs are not usually used for direct traceability of SFR.

The conventional view holds that a subset of lGRBs originates from core-collapse events, often accompanied by supernova-like emission signatures \citep{1998Natur.395..670G,2003Natur.423..847H}. Due to their extreme luminosities, lGRBs serve as powerful tools for probing the SFR at high redshifts. 
Previous studies have shown that the lGRB formation rate increases rapidly at $z < 1$ \citep{2004ApJ...609..935Y, 2006ApJ...642..371K, 2012MNRAS.423.2627W}, consistent with the evolution of the cosmic SFR \citep{2006ApJ...651..142H}. 
However, \citet{2015ApJS..218...13Y} reported a significant excess in the lGRB rate relative to the SFR at $z < 1$ using a sample of 127 \textit{Swift} lGRBs (see also \citealp{2015ApJ...806...44P}), suggesting that the differential comoving volume $\frac{dV(z)}{dz}$ term might have been overlooked in earlier analyses.

\citet{2019MNRAS.488.5823L} compared the lGRB rate derived from a sample of 376 lGRBs observed between 1991 and 2016 with the cosmic SFR from \citet{2014ARA&A..52..415M}. They found that the lGRB rate mildly exceeds the SFR in the redshift range $1 < z < 5$, suggesting that this discrepancy may reflect differences in the evolutionary pathways or environmental dependencies of the GRB progenitor population.
More recently, \citet{2022MNRAS.513.1078D} re-evaluated the lGRB rate using a complete sample and confirmed that the excess at low redshift persists. 
\citet{2024ApJ...963L..12P} further proposed that approximately $60\% \pm 5\%$ of lGRBs at $z < 2$ may originate from compact star mergers, while the remaining $40\% \pm 5\%$ are linked to collapsars, some of which are associated with supernovae.

Recent observations have revealed evidence that some lGRBs are associated with kilonovae \citep{2015NatCo...6.7323Y,2021arXiv210907694J,2022Natur.612..232Y}, while some sGRBs are linked to supernovae \citep{2021NatAs...5..911Z}. These findings challenge the traditional $T_{90}$ classification scheme and suggest that burst duration alone is insufficient to determine the progenitor type. It also shows that neither sGRBs nor lGRBs can be used for trace SFR. The key scientific question becomes whether core collapse GRBs can be used for retrospective SFR.

Given these insights, one would expect SN/GRBs — being directly connected to the death of massive stars — to follow the cosmic SFR more closely than the general GRB population. However, our results indicate that even within this well-defined subsample, a significant excess over the SFR remains at low redshifts ($z < 1$). This suggests several possibilities:
(a) The current sample size and completeness may be insufficient to determine whether GRBs can be used to trace the SFR at low redshifts; (b) GRBs may not be traceable SFR in low redshift.

Potential explanations for this discrepancy include the small size of the current SN/GRB sample, as such events constitute less than $2\%$ of all observed GRBs. The current sample is deficient in high-redshift ($z>1$) SN/GRBs, with only three detected events in the $1 < z < 2$ range, leading to considerable uncertainty. However, future observations may allow us to examine whether the SN/GRB FR follows the cosmic SFR at higher redshifts. 
	
Host galaxy properties may also play a crucial role. Our sample includes only one sGRB, while studies of lGRB host galaxies reveal systematic differences compared to typical star-forming galaxies. These differences include lower stellar masses \citep{2015A&A...581A.102V}, reduced star formation rates \citep{2016A&A...590A.129J}, and lower mass and metallicity. \citep{2019A&A...623A..26P}. The strong metallicity dependence of GRB formation likely contributes to the observed deviation between the GRB rate and the cosmic SFR, particularly at low redshifts. Therefore, the complex environments and progenitor conditions suggest that GRBs at low redshifts might not be reliable tracers of the SFR.

This work was supported by the Postdoctoral Fellowship Program of CPSF under Grant Number GZC20252095, the China Postdoctoral Science Foundation under Grant Number 2025M773194, National Key R\&D Program of China (grant No. 2022YFE0116800), Yunnan Fundamental Research Projects (grant NOs. 202501AS070055, 202503AP140013 and 202401AT070143), the China Manned Space Program (grant No. CMS-CSST-2025-A16), the International Partership Program of Chinese Academy of Sciences (No. 020GJHZ2023030GC) and the National Natural Science Foundation of China (grant Nos. 11933008 and 12303040).

\bibliographystyle{apj}
\bibliography{ms}

\clearpage

\setlength{\tabcolsep}{1mm}{
\renewcommand\arraystretch{1}
\begin{center}
\begin{longtable}{lccccccccc}
\caption{58 SN/GRBs Included in Our analysis}
\label{tab:1} \\
\hline%
GRB &${T_{90}}$  & z & $\alpha$  &$\beta$  &  $\textit{E}_{\rm p}$   & ${E_{\min}}$-${E_{\max}}$ &F  & $L_p$ &ref \\
      &   (s)         &    &  &  &(keV)      & (keV)   &  (ph/cm$^{2}$/s)   & $erg\, s^{-1}$ &    \\
   (1)&(2)& (3) &(4) & (5) &(6) & (7)&(8)&(9)&(10)\\
\hline%
\endhead%
\hline%
\endfoot%
\hline%
\endlastfoot%
970228&56.00 &0.6950 &-1.92 &-&111.48 &40-700&$2.59 ^{+0.16}_{-0.16}\times 10 ^{-6}$$^a$&$1.42^{+0.09}_{-0.09} \times 10 ^{+52}$&\cite{2023MNRAS.524.1096L}\\
980326&9.00 &0.9000 &-1.85 &-&97.25 &40-700&$3.14 ^{+0.32}_{-0.32}\times 10 ^{-7}$$^a$&$3.00^{+0.31}_{-0.31} \times 10 ^{+51}$&\cite{2023MNRAS.524.1096L}\\
980425&31.00 &0.0090 &-1.90 &-&67.89 &40-700&$1.71 ^{+0.27}_{-0.27}\times 10 ^{-7}$$^a$&$7.39^{+1.17}_{-1.17} \times 10 ^{+46}$&\cite{2023MNRAS.524.1096L}\\
990712&30.00 &0.4330 &-1.36 &-&120.64 &40-700&$5.72 ^{+0.48}_{-0.48}\times 10 ^{-6}$$^a$&$6.20^{+0.52}_{-0.52} \times 10 ^{+51}$&\cite{2023MNRAS.524.1096L}\\
991208&60.00 &0.7060 &-1.10 &-2.20 &190.00 &15-2000&$2.00 ^{+0.20}_{-0.20}\times 10 ^{-5}$$^a$&$5.08^{+0.51}_{-0.51} \times 10 ^{+52}$&\cite{2023MNRAS.524.1096L}\\
000911&500.00 &1.0590 &-0.84 &-2.20 &986.00 &15-8000&$2.00 ^{+0.20}_{-0.20}\times 10 ^{-5}$$^a$&$1.22^{+0.12}_{-0.12} \times 10 ^{+53}$&\cite{2023MNRAS.524.1096L}\\
011121&105.00 &0.3620 &-1.22 &-&222.80 &25-100&$6.59 ^{+0.53}_{-0.53}\times 10 ^{-6}$$^a$&$1.14^{+0.09}_{-0.09} \times 10 ^{+52}$&\cite{2023MNRAS.524.1096L}\\
020305&247.00 &1.9800 &-1.06 &-2.30 &245.10 &25-100&$4.70 ^{+0.47}_{-0.47}\times 10 ^{-7}$$^a$&$1.56^{+0.16}_{-0.16} \times 10 ^{+52}$&\cite{2023MNRAS.524.1096L}\\
020405&40.00 &0.6900 &-1.10 &-1.87 &364.00 &15-2000&$5.00 ^{+0.20}_{-0.20}\times 10 ^{-6}$$^a$&$1.17^{+0.05}_{-0.05} \times 10 ^{+52}$&\cite{2023MNRAS.524.1096L}\\
020903&13.00 &0.2500 &-&-2.60 &5.00 &30-400&$2.80 ^{+0.70}_{-0.70}$&$3.93^{+0.98}_{-0.98} \times 10 ^{+48}$&\cite{2023MNRAS.524.1096L}\\
021211&8.00 &1.0100 &-0.81 &-2.37 &46.50 &30-400&$30.00 ^{+2.00}_{-2.00}$&$2.80^{+0.19}_{-0.19} \times 10 ^{+52}$&\cite{2023MNRAS.524.1096L}\\
030329&63.00 &0.1700 &-1.26 &-2.28 &68.00 &30-400&$451.00 ^{+25.00}_{-25.00}$&$6.40^{+0.35}_{-0.35} \times 10 ^{+51}$&\cite{2023MNRAS.524.1096L}\\
030723&31.00 &0.5400 &-&-1.90 &8.90 &30-400&$2.10 ^{+0.40}_{-0.40}$&$1.98^{+0.38}_{-0.38} \times 10 ^{+49}$&\cite{2023MNRAS.524.1096L}\\
031203&30.00 &0.1050 &-1.63 &-&-&50-300&$1.30 ^{+0.13}_{-0.13}$&$4.89^{+0.49}_{-0.49} \times 10 ^{+49}$&\cite{2023MNRAS.524.1096L}\\
040924&5.00 &0.8590 &-1.03 &-&41.10 &20-500&$3.30 ^{+0.35}_{-0.35}\times 10 ^{-6}$$^a$&$1.92^{+0.20}_{-0.20} \times 10 ^{+52}$&\cite{2023MNRAS.524.1096L}\\
041006&25.00 &0.7160 &-1.30 &-&47.70 &30-400&$8.40 ^{+0.30}_{-0.30}$&$3.90^{+0.14}_{-0.14} \times 10 ^{+51}$&\cite{2023MNRAS.524.1096L}\\
050416A&3.00 &0.65 &-1.00 &-3.40 &26.00 &15 -150 &$5.00 ^{+0.50}_{-0.50}$&$7.58^{+0.76}_{-0.76} \times 10 ^{+50}$&\cite{2012MNRAS.421.1256N}\\
050525&9.00 &0.61 &-1.00 &&79.00 &15 -150 &$47.70 ^{+1.00}_{-1.00}$&$7.58^{+0.16}_{-0.16} \times 10 ^{+51}$&\cite{2005GCN..3479....1C}\\
050824&23.00 &0.8300 &-1.00 &-&80.00 &15-150&$0.50 ^{+0.20}_{-0.20}$&$1.73^{+0.69}_{-0.69} \times 10 ^{+50}$&\cite{2023MNRAS.524.1096L}\\
060218&128.00 &0.0330 &-1.65 &-&25.30 &15-150&$0.16$&$4.62\times 10 ^{+46}$&\cite{2023MNRAS.524.1096L}\\
060729&115.30 &0.5400 &-1.60 &-&201.16 &15-150&$1.40 ^{+0.20}_{-0.20}$&$2.84^{+0.41}_{-0.41} \times 10 ^{+50}$&\cite{2023MNRAS.524.1096L}\\
060904B&171.50 &0.7030 &-1.23 &-&80.09 &15-150&$2.50 ^{+0.20}_{-0.20}$&$6.11^{+0.49}_{-0.49} \times 10 ^{+50}$&\cite{2023MNRAS.524.1096L}\\
071112C&15.00 &0.8230 &-&-&-&-&-&-&\cite{2023MNRAS.524.1096L}\\
080319B&125.00 &0.937&-0.82 &-3.87 &651.00 &20 -7000 &$2.17 ^{+0.21}_{-0.21}\times 10 ^{-5}$$^a$&$9.85^{+0.95}_{-0.95} \times 10 ^{+52}$&\cite{2008GCN..7482....1G}\\
081007&10.00 &0.53 &-1.40 &    -                 &40.00 &25 -900 &$2.20 ^{+0.20}_{-0.20}$&$4.26^{+0.39}_{-0.39} \times 10 ^{+50}$&\cite{2008GCN..8369....1B}\\
090618&113.00 &0.54&-1.42 &&134.00 &15 -150 &$38.80 ^{+0.80}_{-0.80}$&$6.47^{+0.13}_{-0.13} \times 10 ^{+51}$&\cite{2009GCN..9534....1S}\\
091127&7.00 &0.49 &-1.27 &-2.20 &36.00 &8 -1000 &$46.90 ^{+0.90}_{-0.90}$&$1.91^{+0.04}_{-0.04} \times 10 ^{+51}$&\cite{2009GCN.10204....1W}\\
100316D&521.88 &0.0140 &-1.88 &-2.35 &9.62 &15-150&$0.1$&$1.70\times 10 ^{+53}$&\cite{2023MNRAS.524.1096L}\\
100418A&7.00 &0.6200 &-2.16 &-2.06 &187.32 &15-150&$1.00 ^{+0.20}_{-0.20}$&$2.79^{+0.56}_{-0.56} \times 10 ^{+50}$&\cite{2023MNRAS.524.1096L}\\
101219B&34.00 &0.5519&-0.33 &-2.12 &70.00 &10 -1000 &$2.00 ^{+0.20}_{-0.20}$&$1.33^{+0.13}_{-0.13} \times 10 ^{+50}$&\cite{2010GCN.11477....1V}\\
101225A&1088.00 &0.8470 &-0.48 &-&96.65 &15-150&$0.09$&$3.34^{+0.00}_{-0.00} \times 10 ^{+50}$&\cite{2023MNRAS.524.1096L}\\
111209A&810.97 &0.6770 &-1.45 &-&768.79 &15-150&$0.50 ^{+0.10}_{-0.10}$&$3.34^{+0.67}_{-0.67} \times 10 ^{+50}$&\cite{2023MNRAS.524.1096L}\\
111228A&99.84 &0.714&-1.90 &-2.70 &34.00 &10 -1000 &$27.00 ^{+1.00}_{-1.00}$&$5.78^{+0.21}_{-0.21} \times 10 ^{+51}$&\cite{2011GCN.12744....1B}\\
120422A&5.35 &0.2800 &-1.19 &-&97.01 &15-150&$0.60 ^{+0.20}_{-0.20}$&$1.79^{+0.60}_{-0.60} \times 10 ^{+49}$&\cite{2023MNRAS.524.1096L}\\
120714B&159.00 &0.40 &-0.29 &&60.80 &15 -350 &$1.80 ^{+1.13}_{-0.68}\times 10 ^{-8}$$^a$&$1.14^{+0.71}_{-0.43} \times 10 ^{+49}$&\cite{2012GCN.13481....1C}\\
120729A&25.47 &0.8000 &-1.61 &-&25.95 &15-150&$2.90 ^{+0.20}_{-0.20}$&$1.03^{+0.07}_{-0.07} \times 10 ^{+51}$&\cite{2023MNRAS.524.1096L}\\
130215A&65.70 &0.597&-1.00 &-1.60 &155.00 &10 -1000 &$3.50 ^{+0.30}_{-0.30}$&$4.76^{+0.41}_{-0.41} \times 10 ^{+50}$&\cite{2013GCN.14219....1Y}\\
130427A&138.24 &0.34 &-0.79 &-3.06 &830.00 &10 -1000 &$1052.00 ^{+2.00}_{-2.00}$&$1.48^{+0.00}_{-0.00} \times 10 ^{+53}$&\cite{2013GCN.14473....1V}\\
130702A&58.88 &0.1450 &-1.82 &-2.49 &10.43 &10-1000&$16.51 ^{+4.69}_{-4.69}$&$8.66^{+2.46}_{-2.46} \times 10 ^{+49}$&\cite{2023MNRAS.524.1096L}\\
130831A&32.50 &0.4791&-1.51 &-2.80 &67.00 &20 -10000 &$2.50 ^{+0.30}_{-0.30}\times 10 ^{-6}$$^a$&$3.71^{+0.44}_{-0.44} \times 10 ^{+51}$&\cite{2013GCN.15145....1G}\\
140506A&111.10 &0.8890 &-0.9&-2&141.00 &10-1000&$14.20 ^{+0.70}_{-0.70}$&$4.23^{+0.21}_{-0.21} \times 10 ^{+51}$&\cite{2014GCN.16220....1J}\\
140606B&22.78 &0.3840 &-1.24 &-2.19 &554.60 &10-1000&$16.26 ^{+1.35}_{-1.35}$&$1.32^{+0.11}_{-0.11} \times 10 ^{+51}$&\cite{2018PASP..130e4202Z}\\
141004A&9.47 &0.57&-1.30 &-&147.00 &10 -1000 &$18.00 ^{+2.00}_{-2.00}$&$2.45^{+0.27}_{-0.27} \times 10 ^{+51}$&\cite{2014GCN.16900....1P}\\
150518&$>1000$&0.2560 &-&-&-&-&-&-&\cite{2023MNRAS.524.1096L}\\
150818A&123.30 &0.2820 &-1.88 &-&100.00 &15-150&$2.40 ^{+0.30}_{-0.30}$&$1.06^{+0.13}_{-0.13} \times 10 ^{+50}$&\cite{2023MNRAS.524.1096L}\\
161219B&6.94 &0.15 &-1.29 &-&61.88 &15 -150 &$5.30 ^{+0.40}_{-0.40}$&$3.23^{+0.24}_{-0.24} \times 10 ^{+49}$&\cite{2016GCN.20308....1P}\\
171010A&107.30 &0.3285 &-1.11 &-2.22 &154.00 &10-1000&$137.30 ^{+4.41}_{-4.41}$&$4.05^{+0.13}_{-0.13} \times 10 ^{+51}$&\cite{2017GCN.21992....1P}\\
171205A&189.40 &0.0368 &-1.22 &-&295.34 &15-150&$1.00 ^{+0.30}_{-0.30}$&$7.00^{+2.10}_{-2.10} \times 10 ^{+47}$&\cite{2023MNRAS.524.1096L}\\
180720B&108.40 &0.6540 &-1.11 &-2.30&631.00 &10-1000&$125.00 ^{+1.00}_{-1.00}$&$4.90^{+0.04}_{-0.04} \times 10 ^{+52}$&\cite{2018GCN.22981....1R}\\
180728A&8.68 &0.117&-1.54 &-2.46 &79.20 &10 -1000 &$231.00 ^{+1.20}_{-1.20}$&$6.00^{+0.03}_{-0.03} \times 10 ^{+50}$&\cite{2018GCN.23053....1V}\\
181201A&172.00 &0.4500 &-1.25 &-2.73 &152.00 &20-15000&$2.88 ^{+0.33}_{-0.33}\times 10 ^{-5}$$^a$&$2.70^{+0.31}_{-0.31} \times 10 ^{+52}$&\cite{2018GCN.23495....1S}\\
190114C&361.50 &0.42&-1.06 &-3.18 &998.60 &10 -1000 &$246.86 ^{+0.86}_{-0.86}$&$5.24^{+0.02}_{-0.02} \times 10 ^{+52}$&\cite{2019GCN.23707....1H}\\
190829A&58.20 &0.0785&-1.41 &-&130.00 &10 -1000 &$25.6$&$3.78\times 10 ^{+49}$&\cite{2019GCN.25575....1L}\\
200826A&1.14 &0.7480 &-0.26 &-2.40 &88.90 &10-1000&$64.30 ^{+2.10}_{-2.10}$&$1.12^{+0.04}_{-0.04} \times 10 ^{+52}$&\cite{2020GCN.28287....1M}\\
201015A&9.78 &0.4260 &-3.03&-&-&15-150&$1.80 ^{+0.40}_{-0.40}\times 10 ^{-7}$$^a$&$3.05^{+0.68}_{-0.68} \times 10 ^{+51}$&\cite{2020GCN.28658....1M}\\
211023A&80.00 &0.3900 &-1.74&-2.55&92.00 &10-1000&$27.40 ^{+0.70}_{-0.70}$&$1.29^{+0.05}_{-0.05} \times 10 ^{+51}$&\cite{2021GCN.30965....1L}\\
221009A&327.00 &0.15 &-1.70 &-&375.00 &10 -1000 &$2385.00 ^{+3.00}_{-3.00}$&$2.02^{+0.00}_{-0.00} \times 10 ^{+52}$&\cite{2022GCN.32642....1L}\\
230812B&3.00 &0.3600 &-0.8&-2.47&273.00 &10-1000&$740.00 ^{+2.00}_{-2.00}$&$4.46^{+0.01}_{-0.01} \times 10 ^{+52}$&\cite{2023GCN.34391....1R}\\

\end{longtable}
\begin{tablenotes}
\textbf{Note}: References for the spectral parameters ($\textit{E}_{\rm p}$, $\alpha $, $\beta $, peak flux , ${E_{\max}}$, and redshift of GRB 201015A, GRB 221009A and GRB 211023A): \url{ https://www.mpe.mpg.de/~jcg/grbgen.html} ;Besides, These main spectrum parameters of other SN/GRB is extract from \citet{2023MNRAS.524.1096L} and \citet{2015ApJS..218...13Y}. $a$: These peak fluxes are in units of erg cm$^{-2}$  s$^{-1}$.
\end{tablenotes}
\end{center}

\end{document}